\documentclass[11pt]{article}
\usepackage{latexsym} \usepackage{epsf}
\usepackage{a4}
\usepackage{amsfonts}
\makeatletter

\expandafter\let\expandafter\reset@font\csname reset@font\endcsname

\makeatother

\def\be{\begin{equation}}
\def\ee{\end{equation}}
\def\bea{\begin{eqnarray}}
\def\eea{\end{eqnarray}}

\def\one#1{#1^{\raise5pt\hbox{$\scriptstyle\!\!\!\!1$}}\,{}}
\def\two#1{#1^{\raise5pt\hbox{$\scriptstyle\!\!\!\!2$}}\,{}}

\makeatletter
\def\binrel@#1{\begingroup
  \setboxz@h{\thinmuskip0mu
    \medmuskip\m@ne mu\thickmuskip\@ne mu
    \setbox\tw@\hbox{$#1\m@th$}\kern-\wd\tw@
    ${}#1{}\m@th$}%
  \edef\@tempa{\endgroup\let\noexpand\binrel@@
    \ifdim\wdz@<\z@ \mathbin
    \else\ifdim\wdz@>\z@ \mathrel
    \else \relax\fi\fi}%
  \@tempa
}
\let\binrel@@\relax
\def\overset#1#2{\binrel@{#2}%
  \binrel@@{\mathop{\kern\z@#2}\limits^{#1}}}
\def\underset#1#2{\binrel@{#2}%
  \binrel@@{\mathop{\kern\z@#2}\limits_{#1}}}
\makeatother
\newcommand{\p}{^{\prime}}
\newfont{\bbd}{msbm10 scaled\magstep1}

\begin{document}
\hfill NTZ 15/2001

\hfill revised

\begin{center}
{\LARGE \bf Universal $R$ operator \\ with deformed conformal symmetry }

\vspace{2cm}

{\large D. Karakhanyan$^{a,b}$, R. Kirschner$^b$ and  M. Mirumyan$^a$}

\end{center}

\begin{itemize}

\item[$^a$]
Yerevan Physics Institute, \\
Br. Alikhanian st. 2, 375036 Yerevan, Armenia.
\item[$^b$]
Naturwissenschaftlich-Theoretisches Zentrum und Institut f\"ur Theoretische
Physik, Universit\"at Leipzig, \\
Augustusplatz 10, D-04109 Leipzig, Germany
\end{itemize}

\vspace{3cm}
\noindent
{\bf Abstract}

We study the general solution of the Yang-Baxter equation with
deformed $sl(2)$ symmetry.
The universal R operator acting on tensor products
of arbitrary representations is obtained in spectral decomposition and
in integral forms.
The results for  eigenvalues, eigenfunctions and integral
kernel appear as deformations of the ones in the rational case.
They provide a basis for the construction of integrable
quantum systems
generalizing the XXZ spin models to the case of arbitrary not
necessarily finite-dimensional representations on the sites.


\renewcommand{\refname}{References.}
\renewcommand{\thefootnote}{\arabic{footnote}}
\setcounter{footnote}{0}
\setcounter{equation}{0}

\renewcommand{\theequation}{\thesection.\arabic{equation}}

\vspace{1cm}

\section{Introduction}
All known exactly solved many-body models belong to two
large classes: two-dimensional classical statistical
models (Ising model, vertex models \cite{bax}) and
one-dimensional quantum models. The standard example of
one-dimensional quantum model is Heisenberg spin chain
which has been studied in much detail for finite-
dimensional representations. The method of solving
the Schr\"odinger equation for this model was pioneered by
Bethe \cite{bethe}. Later it has been understood that the Bethe
ansatz can be applied to models, in which the scattering
between (quasi-) particles is purely elastic. Mathematically
the condition of elasticity is expressed by Yang-Baxter
equation.

The generalization of the Bethe ansatz solution, the
Quantum Inverse Scattering Method, was proposed by the
 group headed by L. D. Faddeev \cite{fst}.

The simplest $SU(2)$-symmetric spin 1/2 Heisenberg chain
was generalized for the cases of lower spin to the
uniaxial $U(1)$-symmetric systems  \cite{yy} and to
biaxial spin systems \cite{bax} and to the case of higher
spins \cite{bt}.

A construction of the universal R-matrix acting on
arbitrary spin ($\frac n2, \frac m2$) representations
of finite dimensions relies on the fusion procedure
\cite{KRS}.
Based on the result in \cite{KR} the models of homogeneous
periodic XXZ chains of arbitrary spins ($\frac n2 = \frac m2$)
have been studied, working out  Hamiltonians, Bethe ansatz and
thermodynmic properties.

Establishing some first concepts of quantum groups
Jimbo \cite{Jimbo}   derived the universal XXZ R operator in
algebraic terms in the spectral decomposition form.
The deformed $sl(2)$ algebra relations appeared first
in \cite{KR81}.

The methods of quantum groups \cite{Drinfeld,Jimbo,FRT,Woronowicz}
allow a calculation in the framework of the q-deformed universal enveloping
algebra of the loop group $\hat {sl(2)}$. This has been
performed in \cite{KST} where the result is given in an algebraic
form in terms of a series.
The result has been used in \cite{BLZ} in the study of the integrable
structure of conformal field theory.
In \cite{Zhang} a method of constructing trigonometric R operators
in spectral decomposition form has been
developed which is applicable to tensor products of any affinizable
representations of quantum algebras and superalgebras.

The general form of L operators (finite dimensional representations)
intertwined by the XXZ fundamental
representation R matrix for the case of the deformation paramenter being a
root of unity has been given in \cite{BS} explaining the relation to the
chiral Potts model. The universal R operator has been constructed for this
case. 

In recent years high energy scattering in gauge field theories
has been discovered \cite{LevPadua} as a new field of applications of
integrable quantum systems. In a number of cases the leading
contribution to the effective interaction appearing in the
Regge or the Bjorken asymptotics of scattering
is determined by  Hamiltonians of integrable chains.
Chains with a few sites are of interest describing the
contribution of the exchange of the corresponding number of
partons or reggeons. Unlike the mostly studied spin chains
however the quantum spaces corresponding to the sites are
infinite dimensional involving all the momentum states
of the partons (longitudinal momentum, one dimensional)
or reggeons (transverse momentum, two dimensional).
Also the application of the Bethe ansatz technique is not
straightforward  \cite{FK}.

Therefore it is necessary to adapt the formulation
and representation
of the known methods of quantum integrable systems
to the special needs of the new applications.
A number of activities are going in this direction
\cite{Lipatov:1993qn,Korchemsky:1995um,Korchemsky:1996be,Lipatov:1999as,
DeVega:2001pu,Derkachov:2001yn,
Karakhanian:2000gy,Derkachov:2000ze}.
The problems and the viewpoints arising from
high energy scattering motivate investigations
being of general interest.

In this context
it is convenient to represent the quantum state of a site
as a wave function. In the case of partons (Bjorken limit)
it would depend on a one dimensional position variable,
in the case of reggeons on a two dimensional position.
The   symmetry algebra acts in terms of differential operators
on the wave functions classified in the corresponding representations.
Operators are conveniently represented in differential or
integral forms.

Owing to applications to the Bjorken asymptotics of QCD
\cite{Braun:1998id,Derkachov:2001km}
in this way the general rational solution of the Yang-Baxter
equation with $sl(2)$ and with $sl(2|1)$ symmetry have been considered
\cite{Derkachov:2000ne,Derkachov:2001sx}.
The algebra $sl(2)$ represents the one-dimensional conformal symmetry
of the asymptotic QCD interaction. In particular it generates
the M\"obius transformations on the argument of the wave function.
The algebra $sl(2|1)$ represents the superconformal
symmetry playing the analogous role in the Bjorken asymptotics
in {\cal N} = 1 supersymmetric Yang-Mills theory.

A scheme emerges which because of  the physical background
is simple and clear. It is based on the well known methods
\cite{KS80,KR81,TTF,Nankai,LesHouches}
 and results in formulations suitable for the
above mentioned applications.
In the present paper we
treat along this line the universal  R operator in the situation
of deformed $ sl(2) $ symmetry. 
Presently we are not able to
indicate the role which  the deformed conformal symmetry in high
energy scattering might play.

We obtain the action of the R operator on the tensor product of arbitrary
representations of the deformed algebra of lowest weight type in
spectral decomposition form by constructing the wave functions of
lowest weight states in the irreducible decomposition with repect to the
co-product generators and the corresponding eigenvalues.
Furthermore, we obtain this universal R operator in integral form.
Our results provide  a basis for the construction of integrable systems
generalizing the XXZ spin chain models to the case where the quantum spaces
on the sites are arbitrary not necessarily finite-dimensional
representations of
the deformed $sl(2)$ algebra.

We restrict ourselves to generic values of the representation parameter.
The representations considered are the ones constructed from a lowest weight
vector.  We shall not discuss here the peculiarities arising in the case of
the deformation paramenter being a root of unity, where cyclic
representations are essential.

We summarize in Sect. 2 some results on the $sl(2)$ symmetric
rational R operators in the form where representations are
written in terms of wave functions, algebra generators represented
as differential operators and the universal R operator is given
in integral form. In Sect. 3 we recall  necessary facts
on the deformed algebra and consider in detail the functions
appearing in the decomposition of the tensor product represention.
In Sect. 4 we formulate the conditions on the universal R operator
following from YBE,  point out their equivalence to the basic intertwining
property and obtain the eigenvalues determining the spectral decomposition of
the universal R operator.
The integral kernel of the R operator is derived in Sect. 5. In Sect. 6
we study the integrals appearing in the action of the obtained
integral operator on the eigenfunctions.

\section{The undeformed case}

\setcounter{equation}{0}

We summarize some basic relations of the $sl(2)$ symmetric (rational)
solutions of the YBE in a formulation, where the representation spaces are
the ones of polynomials and with the symmetry generators acting there as
differential operators \cite{Derkachov:2000ne,Derkachov:2001sx}.
This sets the convenient perspective on our approach.

The best known example of a solution of the YBE is given by the $4 \times 4$
matrix
\be
\label{R1/2}
R^{(-\frac{1}{2}, -\frac{1}{2})}(u)=\left(
\begin{array}{cccc}
a&&&\\
&b&c&\\
&c&b&\\
&&&a
\end{array}\right)
\ee
where $a=u+\eta,\quad b=u,\quad c=\eta $.  $u$ is the spectral parameter
and $\eta$ is the model parameter which can be set equal to unity.
The relation to $sl(2)$ symmetry becomes evident in the expressions
\be
\label{sigma}
R^{(-\frac{1}{2}, -\frac{1}{2})}(u)= u \hat I + \eta \hat {\cal P}
= \hat L^{(-\frac{1}{2})} (u+\frac{1}{2} ; \eta)
\ee

in terms of the permutation operator of two spin-$\frac{1}{2}$ representation
spaces, $ \hat {\cal P} = \frac{1}{2} (I\otimes I + \sigma^a \otimes \sigma^a)
$, or in terms of the $\ell = -\frac{1}{2} $ representation Lax matrix,
which in general looks like

\be
\label{lax}
 \hat L^{(\ell)} (u) = u I_{2 \times 2} \otimes \hat I + \eta \sigma^a \otimes
\hat S^{(\ell) a} = \left (
\begin{array}{cc}
u + \eta \ S^0 & \eta \ S^- \\
\eta S^+ & u - \eta \ S^0
\end{array}
\right )
\ee
Here  $ S^{(\ell) a} $ are  the $sl(2)$
generators
$$
[S^0 ,S^\pm] \ =\ \pm S^\pm,\qquad [S^+,S^-] \ = \ 2S^0 .
$$
acting in the representation space $V^{(\ell)}$ and
$\hat I$ the identity operator in this space.

Moreover the particular YBE for two  fundamental
representations, $ \ell_1 = \ell_2 = - \frac{1}{2} $,
and one arbitrary, $ \ell_3
= \ell $, with the substitution
$$ R^{(-\frac{1}{2}, \ell )} (u) = \hat L^{(\ell )} (u + \frac{1}{2} )
$$
is equivalent to the commutation relations of $\hat S^{(\ell),a} $.
Another particular representation case of YBE, with general $ \ell_1,
\ell_2$ and with $\ell_3 = - \frac{1}{2} $ can be written as
\be
\label{lybe}
 R^{(\ell_1,\ell_2)}_{12}(u-v) \  L^{(\ell_1)}_1(u)  \
L^{(\ell_2)}_2(v) \
= \  L^{(\ell_2)}_2(v)  \  L^{(\ell_1)}_1(u) \
R^{(\ell_1,\ell_2)}_{12} (u-v).
\ee

and involves the universal $sl(2)$ symmetric R-operator, i.e. the one acting
on the tensor product of arbitrary representations $V^{(\ell_1)} \otimes
V^{(\ell_2)}$ . This relation allows to determine the latter from the given
Lax operators \cite{FRT}.

The generic representation space $V^{(\ell )}$ can be chosen to be the space
of polynomials in one variable $x$. Then the $sl(2)$ generators act as

\bea
\label{generators}
 S^{(\ell)-} =  S^{(0) -},  \ \
 S^{(\ell)  0} = x^{- \ell}  \ S^{(0) 0} \ x^{\ell}, \ \
 S^{(\ell) +} =  x^{- 2 \ell} \  S^{(0) + } \ x^{2 \ell}. \cr
 S^{(0) -} = \partial,
 S^{(0) 0 } = x \partial,
 S^{(0) +} = - x^2 \partial,
\eea

Notice that the information on the representation $\ell$
is mainly carried by the differential operators, whereas the representation
spaces look alike. They are spanned by the monomials $x^m$ and $1$
represents the lowest weight vector annihilated by $S^-$.
 However, a symmetric scalar product
and the corresponding normalization, $< x| \ell, m> = C_{\ell, m } x^m $,
depend unavoidably on the representation $\ell$.

In the particular cases of negative half-integer $\ell, \ \ 2 \ell = -N$, the
rising operator $\hat S^{+} $ has a null vector, $ \hat S^{+} x^N = 0 $, and
$\{ x^m \}_0^N $ spans the finite dimensional irreducible representation,
corresponding to the representation of three-dimensional rotations of
angular momentum $-\ell$.

The tensor product space $V^{(\ell_1, \ell_2)} $ is represented by
polynomials in $x_1$ and $x_2$. The action of the generators is given
by (trivial co-product)

\be
\label{trivial}
S^a = S_1^{(\ell_1) a} \ + \ S_2^{(\ell_2) a}.
\ee

The Clebsch-Gordan decomposition is readily recovered by noticing
that
$$ S^- \ (x_1 - x_2)^n = 0, \ \
S^0 \ (x_1 - x_2)^n = \ ( \ell_1 + \ell_2 + n ) \ (x_1 - x_2)^n
$$
and that the irreducible representations are spanned by
$$ \phi_n^{(m)} (x_1, x_2)  = (S^+)^m (x_1 - x_2 )^n .$$

Now we consider $ R_{12}^{(\ell_1, \ell_2)}$ as an operator acting
on functions of $x_1$ and $x_2$. Separating in (\ref{lybe}) the dependence on
$(u+v)$ and $(u-v)$ we obtain two sets of conditions,

\be
\label{sym}
 R_{12}(u) \ S^a\ =\  S^a  \  R_{12}(u),
\ee
and
\be
\label{Krelation}
 R_{12}(u)\ \hat K(u)\ =\ \hat{\bar K}(u) \  R_{12}(u),
\ee
where
$$
\hat K(u)=\left(
\begin{array}{cc}
\frac u2(S^0_2-S^0_1)+S^0_1S^0_2+S_1^-S_2^+&\frac u2(S_2^--S_1^-)
+S^0_1S_2^--S_1^-S^0_2\\
\frac u2(S_2^+-S_1^+)-S^0_1S_2^++S_1^+S^0_2&\frac u2(S^0_1-S^0_2)
+S^0_1S^0_2+S_1^+S_2^-
\end{array}\right),
$$
and
$$
\hat {\bar K} (u)=\left(
\begin{array}{cc}
\frac u2(S^=_2-S^0_1)+S^0_1S^0_2+S_1^+S_2^-&\frac u2(S_2^--S_1^-)
+S_1^-S^0_2-S^0_1S_2^-\\
\frac u2(S_2^+-S_1^+)+S^0_1S_2^+-S_1^+S^0_2&\frac u2(S^0_1-S^0_2)
+S^0_1S^0_2+S_1^-S_2^+
\end{array}\right).
$$

(\ref{sym}) expresses the $sl(2)$ symmetry. It implies that
$ \phi_n^{(m)} (x_1,x_2) $ are eigenfunctions of $\hat R_{12} (u)$
with the eigenvalues $R_n (u)$ independent on $m$.
The matrices
of operators $\hat K, \hat {\bar K} $ transform covariantly,
\be
\label{covariance}
[\sigma^a-S^a,K(u)]=0,\qquad [\sigma^a+S^a,\bar K(u)]=0.
\ee
Therefore the conditions (\ref{Krelation}) are not all independent and
one  of them is sufficient to determine the eigenvalues $R_n$.
We choose the upper right corner component  condition from
(\ref{Krelation}) appearing
as the simplest in our representation, calculate the action of
the operators on $(x_1 - x_2)^n$ and arrive at the recurrence relation
\be
\label{recr}
R_n(u)=-R_{n-1}(u)\frac{u+n -1 +\ell_1 + \ell_2}{-u+n-1+ \ell_1 + \ell_2},
\ee
with the known result

$$
R_n(u)=(-1)^n R_0(u) \prod_{k=1}^n \frac{u+k-1+\ell_1+\ell_2}
{-u+k-1+\ell_1+\ell_2}.
$$

We would like to represent the $R$ operator acting on polynomial functions
$\psi (x_1,x_2) $ in integral form
\be
\label{intform}
 R_{12} (u)\  \psi (x_1,x_2) = \int dx_{1^{\prime}} dx_{2^{\prime}}
 {\cal R} (x_1,x_2 |x_{1^{\prime}},x_{2^{\prime }})
\psi (x_{1^{\prime}},x_{2^{\prime }})
\ee
The integration in $x_{1^{\prime }}, x_{2^{\prime }}$ is along closed
contours.
The defining relations result in a set of differential equations
for the kernel.  A set of four first order differential equations can be
extracted and shown to imply the other equations. The result for the kernel
is given by
\be
\label{kernelun}
 {\cal R} (x_1,x_2 |x_{1^{\prime}},x_{2^{\prime }}) =
{ (x_1 - x_2)^{\alpha } \ (x_{1^{\prime}} - x_{2^{\prime}})^{\delta}
\over (x_{2^{\prime }} - x_1)^{\beta} \ ( x_2 - x_{1^{\prime
}})^{\gamma} },
\ee
$$
\alpha = u - \ell_1 - \ell_2 +1, \ \beta = u + \ell_1 - \ell_2 +1,
\gamma = u - \ell_1 + \ell_2 +1, \ \delta = u + \ell_1 + \ell_2 -1.
$$

The integral representation of the Euler Beta function involving the closed
double-loop Pochhammer contour used for formulating
its analytic continuation shows how the contours in (\ref{intform} )
 are to be chosen.

\section{The deformed algebra}

\setcounter{equation}{0}

We would like to mention the two relations out of the underlying algebraic
structure which will be important in the following. We use the notation
$ [a] = {q^a - q^{-a} \over q - q^{-1} }$.
Here we suppose that the deformation parameter takes generic values,
because  new features  appear in the case $|q|=1$ and
especially if $q$ is a root of unity. 

(1)  The deformed commutation relations are
\be
\label{alg}
[S^0,S^\pm]=\pm S^\pm,\qquad
[S^+,S^-]=\frac{q^{2S^0}-q^{-2S^0}}{q-q^{-1}} = [2 S^0],
\ee
and the corresponding Casimir element is
\be
\label{co}
{{C}}=S^+S^-+[S^0] [S^0-1] =S^-S^++[S^0] [S^0+1].
\ee

(2)  In order to preserve these deformed relations in the action on
tensor products $V_1 \otimes V_2$ the operators $S_1^a$ and $S_2^a$ acting
on $V_1$ and $V_2$ have to be combined in the deformed co-product. There are
two versions of the co-product
 transforming into each other by the  replacement $q \leftrightarrow
q^{-1}$,
$$ \Delta (S^a) = S_{12}^a \ \ \ {\rm and}\ \ \
 \bar \Delta (S^a ) = \bar S_{12}^a, $$
where
\be
\label{coprod}
S^+_{12} =S^+_1q^{S^0_2}+ q^{-S^0_1}S^+_2,
\quad\qquad S^0 _{12} =S^0_1+S^0_2,\qquad\qquad
S_{12}^- = S^-_1 q^{S^0_2} + q^{-S^0_1} S^-_2,
\ee

Representations can be given by defining the action of the above generators
on polynomials of the variable $x$ as

\be
\label{repres}
\hat S^{(\ell) -} = \hat S^{(0) -}, \ \ \
\hat S^{(\ell) 0} = x^{-\ell } \ \hat S^{(0) 0} \ x^{\ell}, \ \ \
\hat S^{(\ell) +} = x^{-2\ell} \ \hat S^{(0) +} \ x^{2 \ell },
\ee
where
\be
\label{repres0}
S^{(0) -} =  D_q = \frac 1x [x\partial],
 \quad
S^{(0) 0} = x\partial, \quad
S^{(0) +}  = - x^2 D_q.
\ee
The monomials $x^m$ span a basis and they are eigenvectors of $S^{(\ell) 0}$ with
eigenvalues $\ell + m$ and    eigenvectors of the Casimir operator
(\ref{co}) with
eigenvalue  $[\ell] [\ell -1]$.

The tensor product $V^{(\ell_1)} \otimes V^{(\ell_2)} $ is represented again
by polynomials in $x_1$ and $x_2$. The action of the algebra and the
decomposition of the product representation into irreducible ones is
determined by the co-product (\ref{coprod}) The subspaces, invariant under
$S^a_{12} $ are spanned by $\varphi_n^{(m)} (x_1, x_2|q) = (S^+_{12})^m
\cdot \varphi_n(x_1, x_2 | q) $.
Here $n$ labels the invariant subspace and $m$ labels the basis spanning
this subspace.
  The lowest weight states,
obeying
\be
\label{lowestweight}
S_{12}^{(0)} \varphi_n(x_1, x_2;q) = (\ell_1 + \ell_2 +n)
\varphi_n(x_1, x_2;q),  \ \
 S_{12}^- \varphi_n(x_1, x_2;q) = 0,
\ee
are given explicitely by
\be
\label{phin}
\varphi_n (x_1,x_2|q) =
\varphi_n (x_1,x_2;\ell_1,\ell_2;q) \ = \ \prod_{k=1}^n (q^{-\ell_1 +1
-k} \ x_1 - q^{k - 1 + \ell_2}  \ x_2 )
\ee
We shall suppress usually the arguments $\ell_1, \ell_2$ if it is clear
to which representations the variables $x_1, x_2$ refer.

The corresponding Casimir operator has the form

\bea
\label{qCasimir}
 C_{12} = q^{S^0_2-S^0_1+1}S_1^+S^-_2+q^{S^0_2- S^0_1-1}
S_1^-S_2^+-(q-q^{-1})^{-2}\left((q+q^{-1})\left(1+q^{2S^0_2-2S^0_1}\right)
\right)+ \cr
q^{2S^0_2}[\ell_1] [\ell_1 - 1] + q^{-2S^0_1} [\ell_2] [\ell_2 - 1]
\eea

and we have the eigenvalue relation

$$
{\hat {C}_{12} } \ \varphi^{(m)}_n(x_1,x_2| q) =[n+\ell_1 + \ell_2 ]
[n+\ell_1+\ell_2-1] \varphi^{(m)}_n(x_1,x_2|q).
$$

For the co-product $ \bar \Delta $ the basis of the invariant subspaces is
given by
$$ \bar \varphi_n^{(m)} (x_1, x_2| q) = \varphi_n^{(m)} (x_1, x_2| q^{-1} ) $$
We have the analogous relations for the co-product operators $\bar S_{12}^a
, \bar C_{12} $ where in all explicite expressions the replacement $q
\leftrightarrow q^{-1}$ has to be done.

The definition of the co-product allows an extension involving an
additional parameter $u$.

$$ \Delta_u (S^a) = S_{12, u}^a,
$$
\bea
\label{ucoprod}
S_{12, u}^{\pm} =  q^{\mp \frac{u}{2} +S^0_2 } S_1^{\pm} \ + \
 q^{\pm \frac{u}{2} - S^0_1 } S_2^{\pm}, \ \ \
S^0_{12, u} \ = \ S^0_1 + S^0_2.
\eea
Now the basis of invariant subspaces is spanned by
$\varphi_n^{(m)} (x_1, x_2 | q,u)$, where
$$ \varphi_n^{(m)} (x_1, x_2| q,u) = (S_{12, u}^+)^m
\varphi_n (x_1, x_2 | q,u) $$

\be
\label{phinu}
 \varphi_n (x_1, x_2 | q,u) \ = \ \varphi_n (q^{-\frac{u}{2} }x_1,
q^{\frac{u}{2}} x_2|  q) = \varphi_n (x_1,x_2; \ell_1+\frac u2, \ell_2 +
\frac u2;q).
\ee
The Casimir operator (\ref{co}) is represented as
\bea
\label{quCasimir}
C_{12, u}  =  q^{S_2^0 -S_1^0} ( q^{2u-1} S_1^- S_2^+ + q^{1-2u} S_1^+
S_2^- ) +
{q +q^{-1} \over (q-q^{-1})^2}
(q^{2S_2^0} + q^{-2S_1^0} - q^{2S_2^0 -2S_1^0 } -1 )  \cr
+ q^{2S_2^0} [\ell_1] [\ell_1-1] + q^{2S_1^0} [\ell_2] [\ell_2-1].
\eea
$ \varphi_n^{(m)} (x_1, x_2 | q,u) $ are its eigenfunctions with
eigenvalues $[\ell_1 + \ell_2 +n] [\ell_1 + \ell_2 +n -1]$.

Correspondingly, the co-product $\bar \Delta_u $ and the basis polynomials
$\bar \varphi_n^{(m)} (x_1, x_2| q,u)$ are defined by the relations
obtained from the above ones by the replacement
$q \leftrightarrow q^{-1}$. This simple remark turns out to be essential in
the following.

The functions $\varphi_n (x_1,x_2; \ell_1,\ell_2;q)$ represent the
lowest weight states and in this sense they are the appropriate  deformations
of $\phi_n (x_1,x_2) = (x_1 - x_2)^n $. The way of deformation depends
on $\ell_1, \ell_2$ and on the co-product.

The conditions from which we have determined the wave functions
representing $V^{(\ell)}$ or $V^{(\ell_1)} \otimes V^{(\ell_2)}$
are difference equations, more precisely difference equations in
$t_i = \ln x_i \ / \ln q $. In particular besides of the constant
function 1 any function periodic in $t$ with period 1, i.e. with the
property $f(q^N x) = f(x), N$ integer, could represent the lowest
weight state of $V^{(\ell)}$. The restriction to polynomial dependence
on $x_i$ fixes the wave functions uniquely.

In the undeformed case the generalization of $\phi_n$ to non-integer powers
$\alpha$
plays an important role in particular in constructing the integral
kernel of the $R$ operator. Therefore we discuss the corresponding
generalization of the deformed  function $\varphi_n$.

For the continuation of $\varphi_n$ (\ref{phin}) away from integer values $n$
the latter variable should be released from its role of a limit of the
index range in the product. Therefore we write
\bea
\varphi_n (x_1,x_2;\ell_1,\ell_2;q) =
x_1^n q^{- n \ell_1 - \frac 12 n(n-1) } \prod_{k=1}^n (1 - {x_2 \over x_1}
q^{2 k - 2 + \ell_1 +\ell_2} ) = \cr
(-x_2)^n q^{n \ell_2 + \frac 12 n(n-1) } \prod_{k=1}^n (1 - {x_1 \over
x_2} q^{-2 k + 2 - \ell_1 -\ell_2} ),
\eea
and change the index $k$ to $k^{\prime } = n - k +1$,
\be
\prod_{k=1}^n ( 1 - {x_1 \over x_2} q^{2n - 2k + \ell_1 + \ell_2}) =
\prod_1^{\infty} { (1 - {x_1 \over x_2} q^{2n - 2k+ \ell_1 +\ell_2})
\over
 (1 - {x_1 \over x_2} q^{ - 2k+ \ell_1 +\ell_2}) }.
\ee
and similar for the other form. The continuation to arbitrary "powers"
$\alpha$ is given by
\bea
\label{phialpha}
\varphi_{\alpha} (x_1,x_2,\ell_1,\ell_2;q) \ = \
\left \{ \begin{array}{cc}
x_1^{\alpha} \ q^{ -\alpha \ell_1 - \frac 12 \alpha (\alpha-1)} \
g({x_2 \over x_1} q^{\ell_1 + \ell_2}, \alpha, q^{-2} ), & |q| > 1, \cr
(-x_2)^{\alpha} \ q^{ \alpha \ell_2 + \frac 12 \alpha (\alpha-1)} \
g({x_1 \over x_2} q^{-\ell_1 - \ell_2}, \alpha, q^{2}), & |q| < 1.
\end{array}\right.
\eea
where we define for $|q|< 1$ (compare \cite{Rahman} )
\be
\label{g}
g(z,a,q) = \prod_1^{\infty} \left ( {1 - z q^{-a +k } \over
1- z q^k } \right ) =
{ ( z q^{a+1} ,q)_{\infty} \over (zq, q)_{\infty} }.
\ee

$\varphi_{\alpha} $ is homogeneous of degree $\alpha$ in the variables
$x_1, x_2$. For dilatations by factors of $q$ or $q^{-1}$ we
have further homogeneity properties,
\bea
\label{homophi}
\varphi_{\alpha} (q x_1,q^{-1} x_2;\ell_1,\ell_2;q) =
\left ( {x_1 q^{-\ell_1 +1} - x_2 q^{\ell_2 -1} \over
x_1 q^{-\ell_1 +1 - \alpha} - x_2 q^{\ell_2 -1 +\alpha} } \right )
\varphi_{\alpha} ( x_1, x_2;\ell_1,\ell_2;q),
\cr
\varphi_{\alpha} (q^{-1} x_1,q x_2;\ell_1,\ell_2;q) =
\left ( {x_1 q^{-\ell_1 -\alpha} - x_2 q^{\ell_2 +\alpha} \over
x_1 q^{-\ell_1 } - x_2 q^{\ell_2 } } \right )
\varphi_{\alpha} ( x_1, x_2;\ell_1,\ell_2;q).
\eea
The multiplication rule holds,
\be
\label{multipl}
\varphi_{\alpha} ( x_1, x_2;\ell_1,\ell_2;q) \
\varphi_{\beta} ( x_1, x_2;\ell_1+\alpha ,\ell_2+ \alpha;q)
=
\varphi_{\alpha+\beta} ( x_1, x_2;\ell_1,\ell_2;q),
\ee
related to the the corresponding property of the undeformed
$\phi_{\alpha} = (x_1 - x_2)^{\alpha} $.
For $\beta = -\alpha$ it results in the inversion rule,
\be
\label{inversion}
(\varphi_{\alpha} ( x_1, x_2;\ell_1,\ell_2;q) )^{-1} =
\varphi_{-\alpha} ( x_1, x_2;\ell_1+\alpha ,\ell_2+ \alpha;q).
\ee
The relations are easily seen to hold for integer $\alpha = n$
(\ref{phin}) and to continue to arbitrary values.
The following relation for integer values
\be
\label{crossingn}
\varphi_{n} ( x_2, x_1;\ell_2,\ell_1;q) ) = (-1)^n
\varphi_{n} ( x_1, x_2;\ell_1,\ell_2;q^{-1}).
\ee
implies the crossing rule
\be
\label{crossing}
\varphi_{\alpha} ( x_2, x_1;\ell_2,\ell_1;q) ) = C_{\alpha}
\varphi_{\alpha} ( x_1, x_2;\ell_1,\ell_2;q^{-1}),
\ee
where $C_{\alpha}$ does not depend on $x_1, x_2$.
We notice that the action  of the co-product generators
$S_{12}^{0}, S_{12}^{-} $ (\ref{lowestweight}) continues to arbitrary
$\alpha$ as well. Finally we write the action of the finite difference
operators,
\bea
\label{diff}
D_{q,1} \varphi_{\alpha} ( x_1, x_2;\ell_1,\ell_2;q) =
[\alpha ] \ q^{-\ell_1 +\frac 12 - \frac{\alpha}{2} }
\varphi_{\alpha-1} ( q^{- \frac 12} x_1, x_2;\ell_1,\ell_2;q), \cr
D_{q,2} \varphi_{\alpha} ( x_1, x_2;\ell_1,\ell_2;q) =
- [\alpha ] \ q^{\ell_2 -\frac 12 + \frac{\alpha}{2} }
\varphi_{\alpha-1} ( q^{- \frac 12} x_1, x_2;\ell_1,\ell_2;q).
\eea

\section{The universal $R$ operator}

\setcounter{equation}{0}

The $4\times 4$ matrix (\ref{R1/2}) with the non-zero elements $a,b,c$
substituted as
\be a= q^{u+1} - q^{-u-1},
b= q^u - q^{-u} ,
c= q - q^{-1}.
\ee
is another well known solution of YBE on which
the anisotropic XXZ Heisenberg chain of spin $\frac{1}{2}$ ,
of higher spins and  related integrable models are based.
The Lax operator
\be
\label{L}
L^{(\ell )}_q (u)=\left(
\begin{array}{cc}
q^{u+S^0}-q^{-u-S^0}&(q-q^{-1})S^-\\
(q-q^{-1})S^+&q^{u-S^0}-q^{-u+S^0}
\end{array}\right).
\ee
involves the generators of the deformed algebra obeying (\ref{alg}).
It represents the   operator $R^{(-\frac{1}{2}, \ell) } (u + \frac{1}{2})$
acting on the tensor product of representation spaces $V^{-\frac{1}{2}}
\otimes V^{(\ell)} $ where the first one is spanned by the two-dimensional
column vectors and the latter by polynomials in $x$.
The YBE for $\ell_1 = \ell_2 =-\frac{1}{2}$ and general $\ell_3 = \ell$ is
equivalent to the deformed commutation relations (\ref{alg}).
The particular YBE for arbitrary $\ell_1, \ell_2$ and for $\ell_3 =
-\frac{1}{2}$ can be formulated as in (\ref{lybe}) in terms of the Lax operator
and the universal  R operator $R^{(\ell_1 \ell_2)}_{12} $
and serves as the defining relation for the
latter. Explicitely we have (the subscripts $(\ell_1,\ell_2)$ and $(12)$
will be suppressed in the following)
\be
\label{ybe}
\!\!\!\!\!R(u-v)
\left(\begin{array}{cc}
q^{u+S^0_1}-q^{-u-S^0_1}&(q-q^{-1})S_1^-\\
(q-q^{-1})S_1^+&q^{u-S^0_1}-q^{-u+S^0_1}
\end{array}\right)\left(\begin{array}{cc}
q^{v+S^0_2}-q^{-v-S^0_2}&(q-q^{-1})S_2^-\\
(q-q^{-1})S_2^+&q^{v-S^0_2}-q^{-v+S^0_2}
\end{array}\right)=
\ee
$$
\!\!\!\!\!\left(\begin{array}{cc}
q^{v+S^0_2}-q^{-v-S^0_2}&(q-q^{-1})S_2^-\\
(q-q^{-1})S_2^+&q^{v-S^0_2}-q^{-v+S^0_2}
\end{array}\right)\left(\begin{array}{cc}
q^{u+S^0_1}-q^{-u-S^0_1}&(q-q^{-1})S_1^-\\
(q-q^{-1})S_1^+&q^{u-S^0_1}-q^{-u+S^0_1}
\end{array}\right) R(u-v),
$$

Separating terms depending only on $u-v$  from the ones
depending also on $u+v$  we arrive at the
following set of equations:

$$
[R (u),q^{S_1+S_2}]=0,
$$
\bea
\label{defrelations}
R(u)\left(q^{\frac u2+S_1}S_2^-+q^{-\frac u2-S_2}S_1^-\right)=
\left(q^{\frac u2-S_1}S_2^-+q^{-\frac u2+S_2}S_1^-\right)R(u),
\cr
R(u)\left(q^{\frac u2-S_1}S_2^++q^{-\frac u2+S_2}S_1^+\right)=
\left(q^{\frac u2+S_1}S_2^++q^{-\frac u2-S_2}S_1^+\right)R(u),
\cr
R(u)\left(q^{\frac u2+S_2}S_1^-+q^{-\frac u2-S_1}S_2^-\right)=
\left(q^{\frac u2-S_2}S_1^-+q^{-\frac u2+S_1}S_2^-\right)R(u),
\cr
R(u)\left(q^{\frac u2-S_2}S_1^++q^{-\frac u2+S_1}S_2^+\right)=
\left(q^{\frac u2+S_2}S_1^++q^{-\frac u2-S_1}S_2^+\right)R(u),
\eea

$$
R(u)\left(q^{u+S_1-S_2}+
q^{-u+S_2-S_1}-(q-q^{-1})^2S_1^-S_2^+\right)=
$$
\be
\label{defc}
\left(q^{u+S_1-S_2}+q^{-u+S_2-S_1}
-(q-q^{-1})^2S_1^+S_2^-\right)R(u),
\ee
$$
R(u)\left(q^{u+S_2-S_1}+q^{-u+S_1-S_2}-
(q-q^{-1})^2S_1^+S_2^-\right)=
$$
$$
\left(q^{u+S_2-S_1}+q^{-u+S_1-S_2}
-(q-q^{-1})^2S_1^-S_2^+\right)R(u).
$$

It is remarkable that   these  relations can be written in
terms of the $u$-dependent deformed generators $S^a_{12, \pm u} $ and
$ \bar S^a_{12, \pm u} $ and their corresponding Casimir operators.

$$
R(u) \  \bar  S_u^a  =  S_{-u}^a  \ R(u), $$
$$
R(u)S_u^a = \bar S_{-u}^a  R(u), $$
\bea
\label{rsrelations}
{ \hat {C}}_{-u}R(u) = R(u){\hat {\bar C}}_u,\qquad\quad
{\hat {\bar C}}_{-u}R(u) =  R(u)  {\hat {C}}_u,
\eea
In terms of the co-product symbols (\ref{coprod}) this can be summarized as
\be
\label{rdelta}
R(u) \ \Delta_u \ = \ \bar \Delta_{-u} \ R(u) , \ \ \
R(u) \ \bar \Delta_u \ = \ \Delta_{-u} \ R(u),
\ee
expressing the basic intertwining property of the $R$ operator
\cite{Drinfeld}.
In this way the meaning of the deformed $sl(2)$ symmetry of the $R$ operator
becomes evident: The different actions on and irreducible decompositions of
the tensor product $V^{(\ell_1)} \otimes V^{(\ell_2)} $ defined according to
the different $u$-dependent co-products $\Delta_u$ and $\bar \Delta_u$ are
mapped by $R(u)$ isomorphically with repect to the algebra (\ref{alg})
to the ones defined by  $\bar \Delta_{-u}$ and $\Delta_{-u}$.

We do have explicitely the irreducible
decompositions of $V^{(\ell_1)} \otimes V^{(\ell_2)} $
with repect to $ \Delta_{\pm u}$ and $\bar \Delta_{\pm u}$ in terms of the
basis polynomials $ \varphi_n^{(m)} (x_1,x_2 | q,u) $
and $ \bar \varphi_n^{(m)} (x_1,x_2 | q,u) $.
The intertwining property implies the eigenvalue relations
\bea
\label{eigenrel}
R(u) \ \varphi_n^{(m)} (x_1, x_2| q,u) \ = \ \bar R_n \ \bar \varphi_n^{(m)}
(x_1,x_2| q,-u) \cr
R(u) \ \bar \varphi_n^{(m)} (x_1, x_2| q,u) \ = \  R_n \  \varphi_n^{(m)}
(x_1,x_2| q,-u)
\eea
The general relation (\ref{rdelta}) implies also that applying these
eigenvalue relations to one projection (e.g. $a \hat = - $) of the defing
relations (\ref{rsrelations})) is enough to define
the universal $R$ operator in terms of the eigenvalues $R_n(u) , \bar R_n(u) $.

To formulate this explicitely from the perspective of section 2 let us
call
the first set of 3 relations in (\ref{rsrelations}) symmetry relations and call
the transformations  $ S_{-u}^a$ left symmetry and $\bar S_u^a$ right
symmetry of $R(u)$. The remaining relations can be formulated as
(compare (\ref{Krelation}) )
\bea
\label{eigenprob}
R(u)
\left(
\begin{array}{cc}
 K^{+-}(u) &  K^-(u) \\
 K^+(u) & K^{-+}(u)
\end{array}
\right)
=
\left(
\begin{array}{cc}
\bar  K^{+-}(u) &  \bar K^-(u) \\
\bar K^+(u) & \bar K^{-+}(u)
\end{array}
\right)
 R(u)
\eea
where
$$
K^-=S_u^-,\quad \bar K^-=\bar S_{-u}^-,\quad
K^+=S_u^+,\quad\bar K^+=\bar S_{-u}^+,
$$
$$
K^{+-}=q^{u+S_1-S_2}+q^{-u+S_2-S_1}-(q-q^{-1})^2S^-_1S_2^+,
$$
$$
\bar K^{+-}=q^{u+S_1-S_2}+q^{-u+S_2-S_1}-(q-q^{-1})^2S^+_1S_2^-,
$$
$$
K^{-+}=q^{u+S_2-S_1}+q^{-u+S_1-S_2}-(q-q^{-1})^2S^+_1S_2^-,
$$
$$
\bar K^{-+}=q^{u+S_2-S_1}+q^{-u+S_1-S_2}-(q-q^{-1})^2S^-_1S_2^+.
$$

The mutual dependence of the relations due to the deformed symmetry is
established by checking the following relations of covariance
by the right symmetry $\bar S_u^a$ for $K^a$ and by the left symmetry
$S_{-u}^a$ for $\bar K^a $,

$$
[\bar S^-_u,K^-]=0=[S^-_{-u},\bar K^-],\qquad
[\bar S^+_u,K^+]=0=[S^+_{-u},\bar K^+],
$$
$$
[\bar S^-_u,K^+]=(q-q^{-1})^{-1}\left(q^{-S_1-S_2}K^{+-}-
q^{S_1+S_2}K^{-+}\right),
$$
$$
[S^-_{-u},\bar K^+]=(q-q^{-1})^{-1}\left(-q^{-S_1-S_2}\bar K^{+-}+
q^{S_1+S_2}\bar K^{-+}\right),
$$
$$
[\bar S^+_u,K^-]=(q-q^{-1})^{-1}\left(-q^{-S_1-S_2}K^{-+}+
q^{S_1+S_2}K^{+-}\right),
$$
$$
[S^-_{-u},\bar K^+]=(q-q^{-1})^{-1}\left(q^{-S_1-S_2}\bar K^{-+}-
q^{S_1+S_2}\bar K^{+-}\right).
$$

We would like to calculate the action of the of both sides of the
$a \hat = -$ components of the relations (\ref{rsrelations}) to
 the polynomials representing
the lowest weight  vectors $\varphi_n(x_1, x_2|q^{\pm 1},\pm u)$.
The operator $S_u^-$ acts on them as
$$
S_u^-\left(
\begin{array}{c}
\varphi_n(x_1,x_2|q,u)\\
\bar\varphi_n(x_1,x_2|q,u) \\
\varphi_n(x_1,x_2|q,-u)\\
\bar\varphi_n(x_1,x_2|q,-u)
\end{array}\right)=  $$
\be
\label{s-action}
\left(
\begin{array}{c}
0\\
(q-q^{-1})[n] [u+n-1-\ell_1-\ell_2]
\bar\varphi_{n-1}(x_1,x_2|q,-u) \\
(q-q^{-1})[u] [n] q^{\ell_2-\ell_1} \varphi_{n-1}(q^{-1}x_1,qx_2|q,-u)\\
(q-q^{-1})[n][n-1+\ell_1+\ell_2]
\bar\varphi_{n-1}(x_1,x_2|q,-u)
\end{array}
\right)
\ee
More relations are obtained by the replacements $q \leftrightarrow q^{-1}$
or $u \leftrightarrow -u$. In fact we use now the first two relations only.
The relations hold also if continuing from integer $n$ to arbitrary
$\alpha$.

Applying now the $a \hat = -$ component of the second relation in (\ref{rsrelations}) on
$\varphi_n (x_1, x_2| q,u) $ we obtain zero on both sides and applying on
$\bar \varphi_n (x_1, x_2| q,u) $ we obtain the recurrence relation for the
eigenvalues $R_n$,
\be
\label{recrel}
R_n=-R_{n-1}
\frac{[u+n-1+\ell_1+\ell_2]}{[-u+n-1+\ell_1+\ell_2]}
\ee
Acting in the analogous way with the $a \hat = -$ component of the
first relation in (\ref{rsrelations}) we obtain that the same recurrence relation
is obeyed by $\bar R_n$. Since $\varphi_0 (x_1, x_2|q,u)  =
\bar \varphi_0 (x_1, x_2|q,u) = 1$ we have $R_0 = \bar R_0 $ and therefore
\be
\label{Rn}
R_{n}=\bar R_{n}=(-1)^n R_0\prod_{k=1}^n
\frac{[u+k-1+\ell_1+\ell_2]}{[-u+k-1+\ell_1+\ell_2]}.
\ee
In different form the spectral decomposition  was obtained first in
\cite{Jimbo}.

\section{The integral kernel of  the $R$ operator}

\setcounter{equation}{0}

We would like to represent the $R$ operator acting of polynomial functions
$\psi (x_1,x_2) $ in integral form (\ref{intform}).
The  integration is along
closed contours in both $
x_{1^{\prime}}$ and $x_{2^{\prime }} $ in order to allow partial
integrations without boundary terms.
The defining conditions (\ref{defrelations}, \ref{defc},
\ref{rsrelations}) of the form
\be  Q \ R(u)  \ = \ R(u) \ \tilde Q,
\ee
where $Q$ and $\tilde Q$ are composed out of $S^a_1 $ and $S^a_2$,
result in conditions on the kernel of the form
\be
\label{defkernel}
\left ( Q_x \ - \ \tilde Q^T_{x^{\prime }} \right ) \
 {\cal R} (x_1,x_2 |x_{1^{\prime}},x_{2^{\prime }})
\ = \ 0.
\ee
$Q_x$ acts on $x_1, x_2$ and $\tilde Q_{x^{\prime }}$ on
$x_{1^{\prime}}$ and $x_{2^{\prime }} $. $ \tilde Q^T$ is obtained
from $\tilde Q$ by partial integration. This conjugation acts on the
generators (\ref{repres}, \ref{repres0}) as
\be
\label{conjugation}
(S^{a (\ell)}_1 )^T \ = \ - \ S^{a (1-\ell )}_1, \ \ \
(S^{a (\ell)}_2 )^T \ = \ - \ S^{a (1-\ell )}_2.
\ee
We introduce the temporary notations
\bea
\label{AB}
\begin{array}{rr}
A_1 = q^{S_1^0 + S_2^0} \
 {\cal R} (x_1,x_2 |x_{1^{\prime}},x_{2^{\prime }}), \ \ \ &
A_2 = q^{S_1^0 - S_2^0}\
 {\cal R} (x_1,x_2 |x_{1^{\prime}},x_{2^{\prime }}),  \cr
A_3 = q^{- S_1^0 + S_2^0}\
 {\cal R} (x_1,x_2 |x_{1^{\prime}},x_{2^{\prime }}), \  &
A_4 = q^{- S_1^0 - S_2^0} \
 {\cal R} (x_1,x_2 |x_{1^{\prime}},x_{2^{\prime }}),  \cr
B_1 = q^{S_{1^{\prime}}^0 + S_{2^{\prime}}^0}\
 {\cal R} (x_1,x_2 |x_{1^{\prime}},x_{2^{\prime }}), \  &
B_2 = q^{S_{1^{\prime}}^0 +  S_{2^{\prime}}^0}
 {\cal R} (x_1,x_2 |x_{1^{\prime}},x_{2^{\prime }}),  \cr
B_3 = q^{- S_{1^{\prime}}^0 + S_{2^{\prime}}^0} \
 {\cal R} (x_1,x_2 |x_{1^{\prime}},x_{2^{\prime }}), \  &
B_4 = q^{- S_{1^{\prime}}^0 - S_{2^{\prime}}^0} \
 {\cal R} (x_1,x_2 |x_{1^{\prime}},x_{2^{\prime }}).
\end{array}
\eea
The lower subscripts on the generators indicate that they act on
 the variable with the
corresponding subscript  and  that they are taken in the corresponding
representation $\ell_1, \ell_2, \ell_{1^{\prime }} = 1- \ell_1,
\ell_{2^{\prime }} = 1- \ell_2 $.
 The defining conditions (\ref{defrelations} \ref{defc}
\ref{rsrelations}) now appear as 8 linear equations for the unknowns
$A_1,...A_4, B_1,... B_4$. The first relation in (\ref{defrelations})
results in
\be
\label{linAB12}
 A_1 - B_4 =  0, \ \ \ \  A_4 - B_1 = 0.
\ee
The next two equations in (\ref{defrelations}) appear as
\bea
\label{linAB34}
{q^{-\frac u2 - \ell_1 } \over x_1 } A_1 +
({q^{\frac u2 - \ell_2} \over x_2} - {q^{ -\frac u2 + \ell_1} \over
x_1}) A_3  -
{q^{\frac u2  +\ell_2} \over x_2 } A_4 \cr
+ (  (1,2) \leftrightarrow (1^{\prime}, 2^{\prime }) , \ A_i
\leftrightarrow B_i ) \ = \ 0,  \cr
- {q^{\frac u2 + \ell_1 } \  x_1 } A_1 +
({q^{\frac u2 - \ell_1}  x_1} - {q^{ -\frac u2 + \ell_2} \
x_2}) A_3  +
{q^{-\frac u2  -  \ell_2} \ x_2 } A_4  \cr
+ (  (1,2) \leftrightarrow (1^{\prime}, 2^{\prime }) , \ A_i
\leftrightarrow B_i ) \ = \ 0. \cr
\eea
The second line in these equations is obtained from the explicitely
written first line by substituting the subscripts on $x$ and $\ell$
as $1 \leftrightarrow 1^{\prime} , 2 \leftrightarrow 2^{\prime }$
and by replacing $A_i$ by $B_i$. The equations resulting from the next
two in (\ref{defrelations}) have a similar form and can be obtained from
(\ref{linAB34}) by the replacement of the lower subscripts on $x$ and
$\ell$ as $1 \leftrightarrow 2, 1^{\prime} \leftrightarrow 2^{\prime} $ and
by the replacement of the unknowns as $A_3 \rightarrow A_2$ and $ B_1
\rightarrow B_2$.

From (\ref{defc}) we obtain
\bea
\label{linAB78}
-{q^{\frac u2 +\ell_2} \over x_2} (q^{-\frac u2 +\ell_1} \ x_1 -
q^{\frac u2 - \ell_2 } \ x_2 ) \ A_2
+ {q^{-\frac u2 -\ell_2} \over x_2} (q^{-\frac u2 +\ell_2} \ x_2 -
q^{\frac u2 - \ell_1 } \ x_1 ) \ A_3 \cr
+ {x_1 \over x_2} (q^{\ell_1 - \ell_2} \ A_1 \ + \
q^{\ell_2 - \ell_1} \ A_4 )  \cr
-{q^{-\frac u2 +\ell_{1^{\prime }}} \over x_{1^{\prime }} }
(q^{-\frac u2 +\ell_{1^{\prime }} } \ x_{1^{\prime }} - q^{\frac u2 -
\ell_{2^{\prime }} } \ x_{2^{\prime }} )
\ B_2 + {q^{\frac u2 +\ell_{1^{\prime }} } \over x_{1^{\prime }} }
(q^{-\frac u2 +\ell_{2^{\prime }} } \ x_{2^{\prime }}  - q^{\frac u2 -
\ell_{1^{\prime }}  } \ x_{1^{\prime }} ) \ B_3  \cr
- {x_{2^{\prime }} \over
x_{1^{\prime }}} ( q^{\ell_{2^{\prime}} - \ell_{1^{\prime}}} \ B_1 \ + \
q^{\ell_{1^{\prime }}  - \ell_{2^{\prime }} } \ B_4 ) \ = \ 0.
\eea
and the analogous equation obtained from the latter by the substitution
of subscripts on $x$ and $\ell$ $ as 1 \leftrightarrow 2, 1^{\prime }
\leftrightarrow 2^{\prime } $.

We start solving this linear system by excluding $B_i$ and arrive at
\bea
\label{resultA}
 A_2 = { ( q^{-\frac u2 + \ell_1 -1} x_{1^{\prime }} -
q^{\frac u2 +\ell_2 }\  x_2 ) \
  ( q^{-\frac u2 + \ell_1 -1} x_2  -
q^{\frac u2 -\ell_2 +1 }\  x_1 ) \over
 ( q^{-\frac u2 + \ell_1 } x_1  -
q^{\frac u2 -\ell_2 }\  x_2 ) \
  ( q^{-\frac u2 + \ell_1 -1} x_2  -
q^{\frac u2 +\ell_2  }\  x_{1^{\prime }} ) } \ \ A_1, \cr
 A_3 = { ( q^{-\frac u2 + \ell_2 -1} x_1 -
q^{\frac u2 -\ell_1 +1 }\  x_2 ) \
  ( q^{-\frac u2 + \ell_2 -1} x_{2^{\prime }}  -
q^{\frac u2 +\ell_1  }\  x_1 ) \over
 ( q^{-\frac u2 + \ell_2 } x_2  -
q^{\frac u2 -\ell_1 }\  x_1 ) \
  ( q^{-\frac u2 + \ell_2 -1} x_1  -
q^{\frac u2 +\ell_1  }\  x_{2^{\prime }} ) } \ \ A_1, \cr
 A_4 = { ( q^{-\frac u2 + \ell_1 -1} x_{1^{\prime }} -
q^{\frac u2 +\ell_2 }\  x_2 ) \
  ( q^{-\frac u2 - \ell_1 } x_{2^{\prime }}  -
q^{\frac u2 +\ell_2 +1 }\  x_1 ) \over
 ( q^{-\frac u2 - \ell_2 } x_2  -
q^{- \frac u2 -\ell_1 +1 }\  x_{1^{\prime }} ) \
  ( q^{-\frac u2 + \ell_2 -1} x_1  -
q^{\frac u2 +\ell_1  }\  x_{2^{\prime }} ) } \ \ A_1.
\eea
$B_1, B_4 $ are known by (\ref{linAB12}). $B_2, B_3$ are obtained by
excluding $A_i$,
\bea
\label{resultB}
 B_2 = { ( q^{-\frac u2 + \ell_1 -1} x_{1^{\prime }} -
q^{\frac u2 +\ell_2 }\  x_2 ) \
  ( q^{-\frac u2 - \ell_1 } x_{2^{\prime }}  -
q^{\frac u2 +\ell_2  }\  x_{1^{\prime }} ) \over
 ( q^{-\frac u2 + \ell_1 -1} x_2  -
q^{ \frac u2 +\ell_2  }\  x_{1^{\prime }} ) \
  ( q^{-\frac u2 - \ell_1 +1} x_{1^{\prime }}  -
q^{\frac u2 +\ell_2 -1 }\  x_{2^{\prime }} ) } \ \ A_1, \cr
 B_3 = { ( q^{-\frac u2 - \ell_2 } x_{1^{\prime }} -
q^{\frac u2 +\ell_1 }\  x_{2^{\prime }} ) \
  ( q^{-\frac u2 + \ell_2 -1} x_{2^{\prime }}  -
q^{\frac u2 +\ell_1  }\  x_{1} ) \over
 ( q^{-\frac u2 - \ell_2 +1} x_{2^{\prime }}  -
q^{ \frac u2 +\ell_1 -1  }\  x_{1^{\prime }} ) \
  ( q^{-\frac u2 + \ell_2 -1} x_{1}  -
q^{\frac u2 +\ell_1 }\  x_{2^{\prime }} ) } \ \ A_1.
\eea
We substitute $A_i, B_i$ by the definitions in terms of ${\cal R}$
(\ref{AB}) and obtain that the defining conditions result in
four difference equations,

\bea
\label{resultR}
 {\cal R} (q^2 x_1,x_2 |x_{1^{\prime}},x_{2^{\prime }})\!=\!
q^{-2\ell_1}  { ( q^{-\frac u2+ \ell_2 -1} x_2\! -\!
q^{\frac u2 -\ell_1 +1 }x_1 )
  ( q^{-\frac u2 + \ell_2 } x_1\!  -\!
q^{\frac u2 +\ell_1  } x_{2^{\prime }} ) \over
 ( q^{-\frac u2 +\ell_2 } x_1\!  -\!
q^{ \frac u2 -\ell_1  }  x_2 ) \
  ( q^{-\frac u2 + \ell_2 -1} x_{2^{\prime}}  -
q^{\frac u2 +\ell_1 +1  }\  x_1 ) } \cr
 {\cal R} (x_1,x_2 |x_{1^{\prime}},x_{2^{\prime }}), \cr
 {\cal R} ( x_1,q^2 x_2 |x_{1^{\prime}},x_{2^{\prime }}) =
q^{-2\ell_2}  { ( q^{-\frac u2 + \ell_1 -1} x_1 -
q^{\frac u2 -\ell_2 +1 }\  x_2 ) \
  ( q^{-\frac u2 + \ell_1 } x_2  -
q^{\frac u2 +\ell_2  }\  x_{1^{\prime }} ) \over
 ( q^{-\frac u2 + \ell_1 -1 } x_{1^{\prime }}  -
q^{ \frac u2 +\ell_2 +1  }\  x_2 ) \
  ( q^{-\frac u2 + \ell_1 } x_{2}  -
q^{\frac u2 -\ell_2   }\  x_1 ) }
 {\cal R} (x_i), \cr
\! \!  {\cal R} ( x_1,x_2 |q^2 x_{1^{\prime}},x_{2^{\prime }}) =
{q^{-2 +2 \ell_1}   ( q^{-\frac u2 + \ell_1 } x_{1^{\prime }} -
q^{\frac u2 +\ell_2  }\  x_2 ) \
  ( q^{-\frac u2 - \ell_1 } x_{2^{\prime }}  -
q^{\frac u2 +\ell_2  }\  x_{1^{\prime }} ) \over
 ( q^{-\frac u2 + \ell_1 -1} x_2  -
q^{ \frac u2 +\ell_2 +1 }\  x_{1^{\prime }} ) \
  ( q^{-\frac u2 - \ell_1 +1} x_{1^{\prime}}  -
q^{\frac u2 +\ell_2 -1  }\  x_{2^{\prime }} ) }
 {\cal R} (x_i), \cr
\!\!\! {\cal R} ( x_1,x_2 |x_{1^{\prime}},q^2 x_{2^{\prime }}) =
{q^{-2+2\ell_2}   ( q^{-\frac u2 + \ell_2 } x_{2^{\prime }} -
q^{\frac u2 +\ell_1  }\  x_1 ) \
  ( q^{-\frac u2 - \ell_2 } x_{1^{\prime }}  -
q^{\frac u2 +\ell_1  }\  x_{2^{\prime }} ) \over
 ( q^{-\frac u2 + \ell_2 -1} x_1  -
q^{ \frac u2 +\ell_1 +1 }\  x_{2^{\prime }} ) \
  ( q^{-\frac u2 - \ell_2 +1} x_{2^{\prime}}  -
q^{\frac u2 + \ell_1 -1  }\  x_{1^{\prime }} ) }
 {\cal R} (x_i).
\eea
Comparing the dependence on $x_1, x_2, x_{1^{\prime }}, x_{2^{\prime }}$
of the factors appearing on the r.h.s in (\ref{resultR}) with
(\ref{homophi}) leads to the ansatz
\be
\label{ansatzR}
 {\cal R} (x_1,x_2 |x_{1^{\prime}},x_{2^{\prime }}) =
{ \varphi_{\alpha} (x_1,x_2;a_1,a_2;q) \ \varphi_{\delta}
(x_{1^{\prime}}, x_{2^{\prime }}; d_{1^{\prime}}, d_{2^{\prime }};q)
\over
 \varphi_{\beta} (x_{2^{\prime }},x_1;b_{2^{\prime }},b_1;q) \
\varphi_{\gamma}
(x_2, x_{1^{\prime }}; c_2, c_{1^{\prime }};q) }
{\cal R}_0.
\ee
This is indeed a deformation of the kernel found for the undeformed
case(\ref{kernelun}). The detailed comparison of the ansatz with
(\ref{resultR})
results in the following conditions for the parameters,
\bea
\label{parameters}
\alpha = u - \ell_1 - \ell_2 +1, \ a_1 = a - \frac u2 + \ell_1, \ a_2 =
- a - \frac u2 + \ell_2, \cr
\beta = u + \ell_1 - \ell_2 +1, \ b_{2^{\prime }} = b - \frac u2 -
\ell_1, \ b_1 = - b - \frac u2 + \ell_2, \cr
\gamma = u - \ell_1 + \ell_2 +1, \ c_2 = c - \frac u2 - \ell_2, \
c_{1^{\prime }} = - c - \frac u2 + \ell_1, \cr
\delta = u + \ell_1 + \ell_2 -1, \ d_{1^{\prime}} = d - \frac u2 -
\ell_2 +1, \ d_{2^{\prime}}  = - d - \frac u2 - \ell_1 +1.
\eea
$a,b,c,d$ remain unconstrained; they affect the normalization only.
Notice that each equation
of (\ref{resultR}) fixes the parameter triple of two functions in the
ansatz. The underlying YBE guarantees the compatibility of the four
difference equations (\ref{resultR}) and in particular the compatibility
of the conditions on the parameters in our ansatz.

The kernel of the universal $R$ operator is given by (\ref{ansatzR}) with
the parameters substituted as in (\ref{parameters}).

Being the solution of the difference equations (\ref{resultR}) the
integral kernel (\ref{ansatzR}) is not unique. Expressions differing
from the given one by factors being periodic functions of $x_1, x_2,
x_{1\p}, x_{2\p} $ (in the multiplicative sense, $f(q^N x) = f(x), N$
integer ) solve (\ref{resultR}) as well. (\ref{ansatzR}) is the solution
close to the undeformed kernel in its analytic properties.

\section{Deformed Beta integrals}

\setcounter{equation}{0}

Consider the functions $\varphi_{\alpha} (x_1,x_2; \ell_1, \ell_2;q) $
for particular values $\ell_1 = \ell_2 = \frac{1 - \alpha}{2} $ and define
\be
\label{propagator}
\pi_{\alpha} (x_1, x_2) = \varphi_{\alpha} (x_1,x_2; \frac{1 - \alpha}{2},
 \frac{1 - \alpha}{2} ;q) =
\varphi_{\alpha} (x_1,x_2; \frac{1 - \alpha}{2}, \frac{1 -
\alpha}{2};q^{-1}).
\ee
The lowest weight property (\ref{lowestweight}) implies in this case
\be
\label{propsymm}
\left ( S_1^{(\frac{-\alpha}{2}), a} +  S_2^{(\frac{-\alpha}{2}). a}
\right ) \ \pi_{\alpha} (x_1, x_2) \ = \ 0.
\ee
This means that $\pi_{\alpha}(x_1, x_2)$ are the appropriate deformation
of the conformal two-point functions.

(\ref{propsymm}) is checked by writing (\ref{lowestweight})
 explicitely for the case 
$\ell_1 = \ell_2 = -\frac{\alpha}{2} $. The check goes straightforward for
the component $a=0$; explicitely we have
\bea 
\label{S0phi}
\left ( S_1^{\ell_1, 0} + s_2^{\ell_2, 0} - \ell_1 -\ell_2 -\alpha \right ) 
\varphi_{\alpha} (x_1, x_2, \ell_1, \ell_2;q)  \ = \cr
\ ( x_1 \partial_1 + x_2
\partial_2 - \alpha ) \ \varphi_{\alpha} (x_1, x_2, \ell_1, \ell_2;q) 
\ = \ 0. 
\eea 
Then by applying the latter relation $ S_{12}^- \varphi_{\alpha} (x_1, x_2,
-\frac{\alpha}{2}, -\frac{\alpha}{2} ) = 0 $ implies (\ref{propsymm}) for the
component $a=-$. Further, (\ref{S0phi}) leads to 
\be
\label{S-phi}
(S_1^- + S_2^- ) \varphi_{\alpha} (x_1, x_2, \ell_1, \ell_2; q ) = 
(x_1 x_2 )^{-1} \left (S_1^{-\frac{\alpha}{2}, +} + 
S_2^{-\frac{\alpha}{2}, +} \right ) 
\varphi_{\alpha} (x_1, x_2, \ell_1,\ell_2; q ) 
\ee 
and therefore the relation (\ref{propsymm}) for the components $a=0$ and
$a=-$ imply the relation for $a =+$.

We show that
\be
\label{betaint}
\int_{C} \pi_{\alpha-1} (x_1, x\p ) \ \pi_{\beta-1} (x\p, x_2 ) dx\p \ = \
\pi_{\alpha + \beta -1} (x_1, x_2) \ B_C (\alpha, \beta ;q).
\ee
The closed contour $C$ should be choosen such that the integral exists and is
not vanishing identically. The factor independent of $x_1,x_2 $ on the
right-hand side is proportional to the deformed Beta function,
\be
\label{gamma}
B_C (\alpha, \beta;q) = B_C^{0} \ \ { \Gamma_q(\alpha)  \Gamma_q(\beta )
\over \Gamma (\alpha + \beta ) }, \ \ \
\Gamma_q (\alpha + 1) = [\alpha] \ \Gamma_q (\alpha ).
\ee

We check first that the left-hand side of (\ref{betaint})
is annihilated by the
action of $ S_1^{(\frac{1-\alpha-\beta}{2}), a} +
S_2^{(\frac{1-\alpha-\beta}{2}), a} $.
For the component $a = -$ the generators are represented by the finite
difference operators for which integration by parts applies and the
property of $\pi_{\alpha}$ (\ref{propsymm}) implies the corresponding
property for the l.h.s. For the component $a = 0$ the assertion is checked
easily since the dilatation is not deformed. After this the relations
(\ref{S0phi}) and (\ref{S-phi}) 
 imply that also the $a = +$ component annihilates
the l.h.s of (\ref{betaint}). Relying on this we see that the integral has
the above form with some factor $B_C(\alpha, \beta;q)$ independent of $
x_1, x_2 $. By acting with the finite difference operators on both sides
using (\ref{diff}) we obtain iterative relations for $B_C(\alpha, \beta;q)$,
\bea
\label{betaiter}
[\alpha-1]\ B_C(\alpha -1, \beta) \ = \ [\alpha + \beta-1]\
B_C(\alpha,\beta ), \cr
[\beta-1]\ B_C(\alpha, \beta -1) \ = \ [\alpha + \beta-1]\
B_C(\alpha,\beta ).
\eea
In this way we confirm (\ref{gamma}).

(\ref{betaint}) is the appropriate deformation of the classical Beta
integral. The singularities of the integrand are series of poles and
branch points of the type $(x-a)^{\alpha }, \ (x-b)^{\beta} $. The
example of the classical Beta function shows, how closed contours can be
chosen (Pochhammer double-loop contour).

Consider now the action of the integral operator (\ref{ansatzR},
\ref{parameters})
on the lowest weight functions $\varphi_n(x_1, x_2|q,u) $ (\ref{phinu}).
We would like to check by explicite calculations that these functions are
eigenfunctions in the sense of (\ref{eigenrel}) and that the eigenvalues
$R_n$ are the ones obtained previously (\ref{Rn}).
\bea
\label{kernelint}
 R(u) \varphi_n (x_1, x_2|q,u) \ = \
\int_{C_1} dx_{1\p}  \int_{C_2} dx_{2\p}
{\cal R}(x_1,x_2|x_{1\p}, x_{2\p} ) \
\varphi_{n} (x_1,x_2; \frac u2 + \ell_1, \frac u2 + \ell_2;q)
\eea
We substitute the kernel (\ref{ansatzR}, \ref{parameters}) and observe
owing to (\ref{multipl}, \ref{propagator}) that
\bea
\label{phipi}
\varphi_{\delta} (x_{1\p},x_{2\p}; d_{1\p}, d_{2\p};q)
\varphi_{n} (x_{1\p},x_{2\p}; \frac u2 + \ell_1, \frac u2 + \ell_2;q)
\cr  = \
q^{\frac 12 (n - \ell_1 + \ell_2)(u + \ell_1 + \ell_2 + n -1) }
\pi_{u + \ell_1 + \ell_2 +n-1} (q^{-n} x_{1\p}, x_{2\p} ), \cr
(\varphi_{\beta} (x_{2\p},x_{1}; b_{2\p}, b_{1};q) )^{-1}
\ = \
q^{\frac 12 (\ell_1 + \ell_2)(-u-\ell_1+\ell_2-1)} \
\pi_{-u -\ell_1+\ell_2-1} (x_{2\p}, x_1 ), \cr
(\varphi_{\gamma} (x_{2},x_{1\p}; c_{2}, c_{1\p};q) )^{-1}
\ = \
q^{\frac 12 (\ell_1 + \ell_2)(-u+\ell_1-\ell_2-1)} \
\pi_{-u +\ell_1-\ell_2-1} (x_{2}, x_{1\p} ), \cr
\varphi_{\alpha} (x_{1},x_{2}; a_{1}, a_{2};q)
\ = \
q^{\frac 12 (\ell_1 + \ell_2)(u-\ell_1-\ell_2+1)} \
\pi_{u -\ell_1-\ell_2+1} (x_{1}, x_{2} ).
\eea

In (\ref{kernelint}) we have the integral over $x_{2\p}$,
\bea
\label{intx2p}
\int_{C_2} dx_{2\p}
(\varphi_{\beta} (x_{2\p},x_{1}; b_{2\p}, b_{1};q) )^{-1}
\varphi_{\delta} (x_{1\p},x_{2\p}; d_{1\p}, d_{2\p};q)
\varphi_{n} (x_{1\p},x_{2\p}; \frac u2 + \ell_1, \frac u2 + \ell_2;q)
= \cr
q^{-\frac 12 n^2 - \frac n2 (-u +2\ell_2-1) -u \ell_1 (\ell_1 + \ell_2)
(\ell_2 - \ell_1) - \ell_2 } \cr
\ \pi_{2\ell_2 + n-1} (x_{1\p}, q^n x_1 ) \
B_{C_2} (u + \ell_1+\ell_2+n, -u-\ell_1+\ell_2 ).
\eea
Here the first relation in (\ref{phipi}) and (\ref{betaint}) have been
applied.

Then we have the integral over $x_{1\p}$,
\bea
\label{intx1p}
\int_{C_1} dx_{1\p}
(\varphi_{\gamma} (x_{2},x_{1\p}; c_{2}, c_{1\p};q) )^{-1}
\ \pi_{2\ell_2 +n-1} (x_{1\p}, q^n x_1) \cr
 = \
q^{\frac 12 (\ell_1+\ell_2)(-u+\ell_1 -\ell_2-1) }
\ \pi_{-u+\ell_1+\ell_2+n-1} (x_2, q^n x_1) \
B_{C_1} (-u+\ell_1-\ell_2, 2\ell_2 +n ).
\eea
We rewrite the position dependent factor on r.h.s. of (\ref{intx1p})
using the multiplication rule (\ref{multipl}),
\bea
\pi_{-u+\ell_1+\ell_2-1} (x_2, q^n x_1) \ = \
q^{\frac n2 (-u+\ell_1+\ell_2+n-1)} \cr
\varphi_{-u+\ell_1+\ell_2+n-1} (x_2, x_1; 1+\frac n2 -
\frac{\ell_1+\ell_2}{2}, \frac n2 -\frac{\ell_1+\ell_2}{2}; q) \cr
 = \
q^{\frac{n^2}{2} + \frac n2 (-u+2\ell_2-1) } \
\pi_{-u+\ell_1+\ell_2-1} (x_2,x_1) \
\varphi_n (x_2,x_1; -\frac u2 + \ell_2, -\frac u2+\ell_1; q).
\eea
In this way we obtain the result of the action of the integral operator
(\ref{kernelint})
\bea
\int_{C_1} dx_{1\p} \ \int_{C_2} dx_{2\p} \
{\cal R}(x_1,x_2|x_{1\p}, x_{2\p} ) \ \
\varphi_{n} (x_1,x_2; \frac u2 + \ell_1, \frac u2 + \ell_2;q) \cr
\cr  = \
q^{-u (2\ell_1 + \ell_2) + (\ell_1+\ell_2) \ell_2 -\ell_1 -2\ell_2} \
{\cal R}_0 \ \cr
\pi_{u-\ell_1-\ell_2 +1} (x_1, x_2) \ \
\pi_{-u+\ell_1+\ell_2 -1} (x_2, x_1) \
\varphi_n (x_2,x_1; -\frac u2 + \ell_2, -\frac u2+\ell_1; q) \cr
\ \ B_{C_2} (u+\ell_1+\ell_2+n, -u -\ell_1+\ell_2) \ \
B_{C_1} (-u+\ell_1-\ell_2, 2\ell_2 +n)
\cr
\ = \ {\rm const.} \ R_n \ \
\varphi_n (x_1,x_2; -\frac u2 + \ell_1, -\frac u2+\ell_2; q^{-1}).
\eea
The constant depends on $u, \ell_1,\ell_2 $ and of the choice of the
contours but not on $n$. Therefore the
normalization ${\cal R}_0$ of the kernel (\ref{ansatzR}) can be choosen
such that const. $=1 $.
By explicite integration we have shown that the integral form
(\ref{ansatzR}, \ref{parameters}) of the
universal $R$ operator obeys the eigenvalue relation (\ref{eigenrel},
\ref{Rn}).

It may be of interest to check also the Yang-Baxter relation in the
integral representation. For this aim one should try to extend the
corresponding calculations of \cite{Derkachov:2001yn} 
and in particular the
star-triangle relation for integrals over two-point functions.

\section{Discussion}

Our study of YBE with deformed $sl(2)$ symmetry follows the scheme used
earlier \cite{Derkachov:2000ne,Derkachov:2001sx}.
 Generic representations of the algebra of lowest weight
on (one-point) functions provide the starting point.
As the next step the action on
tensor products represented by functions of two points are studied.
Then the YBE involving the given Lax operator and
the universal R operator is considered as a defining relation of the latter.
Evaluating these conditions on the lowest weight two point functions
results in the spectral decomposition form of the universal R operator.
Writing the conditions with R in integral form results in equations for the
kernel.
The scheme relies on the well known standard methods
\cite{KS80,KR81,TTF,Nankai,LesHouches}
reformulated in a way motivated by problems of high energy scattering
and in particular adapted to the treament of non-compact
representations.

In our case the action on tensor products is formulated in terms of the
generators acting on one-point functions by non-trivial co-products.
We introduce the affine extension of the deformed $sl(2)$ co-product
involving the spectral parameter; this replaces the explicite treatment of
the loop algebra.

We devote much attention to the two-point functions representing
the lowest weight states appearing in the irreducible decomposition
of the tensor product representations
with repect to a particular co-product. They play the key role in
formulating the spectral decomposition form of the R operator.
Continuing in their expression the integer representation index to arbitrary
values we obtain the building blocks for the integral kernel.

Fucussing on similarities to the undeformed case we prefer polynomial
functions for describing the representations. The eigenvalue equations from
which these functions are determined allow other solutions differing from
the preferred ones by multiplicatively periodic functions. A formulation in
terms of classes of functions where the equivalence relation is given by this
periodicity may be reasonable.

The defining relations of the universal R operator resulting from the YBE
involving $R(u)$  together with the given Lax operator are shown to be
equivalent to the intertwing property of the operator $R(u)$.
The intertwining property is the statement that the action of
$R(u)$ maps symmetrically the action of the co-products $\Delta_u,
\bar \Delta_u $ to the ones defined correspondingly by $\bar \Delta_{-u},
\Delta_{-u} $. This observation, expected on general grounds
\cite{Drinfeld},
is a further key point in our analysis.

We evaluate the intertwining relations on the two-point wave functions
representing the lowest weight states in the decomposition of the tensor
product with respect to those co-products. We obtain an iterative relation
for the eigenvalues which is easily solved. This result provides the
R operator in the spectral decomposition form.

Further we study the intertwining relations with the R operator written in
integral form. We solve the resulting conditions on the kernel
which appear as 4 difference equations. A particular solution for
the kernel is obtained in terms of the continued lowest weight two-point functions.

A particular case of the latter functions  represent the deformed
analogon of the conformal two-point functions. They are used to formulate a
deformation of the Beta integral. This is another key point which
allows to calculate the action of the integral operator by doing the
integrations and to check the integral form of the R operator against its
spectral decomposition form.

It is remarkable that these integrations are done relying basically on the
symmetry relations. In this way we can avoid the use of extensive summation
formulae \cite{Rahman}.

It is interesting to compare the calculations step by step with the ones in
the undeformed case and to see in particular how the eigenvalues
as functions of the
representation paramenters, the lowest weight and the conformal
two-point functions, the defining relations,  the kernel
of the R operator and the Beta integrals are deformed.

The defining conditions of the undeformed $sl(2)$ symmetric universal
R operator are equivalent to
a system of 4 first order differential equations. In the deformed case
they are replaced by 4 difference equations.

In the undeformed case the defining conditions on the universal R operator
decompose in the symmetry condition and a further one. Operators obeying the
symmetry condition only act symmetrically on the tensor product;
their kernels  are conformal
4-point functions. The R operator is a particular symmetric operator. The
further condition is to fix the arbitrariness left in the conformal 4-point
functions, the dependence on the anharmonic ratio.
Under the deformation the symmetry condition generalizes in two independent
ways into the intertwining relations. They are equivalent to all defining
conditions and fix the R operator completely.

Our results apply to the case of generic values of $q$. The special features
arising for $q$ being a root of unity have not been considered here. Also we
did not investigate the interesting behaviour of the above results in the 
vicinity of such special values of $q$. 
Nevertheless, a similarity between the form of the R operator for the root
of unity case obtained in \cite{BS} and of our integral kernel $ {\cal R} $
can be observed: both are products of four factors. Whereas in \cite{BS}
the factors are Boltzmann weights of the chiral Potts model here they are
two-point functions, continuations of the functions representing the lowest
weight states.

\section*{Acknowledgements}
One of us (DK) is grateful for hospitality at Leipzig University.
The work of two of us (DK and RK) 
has been supported by the German Federal Ministry BMBF/IB.
One of us (MM) has been supported by VW Stiftung.


\begin{thebibliography}{99}
\bibitem{bax} R. J. Baxter 1982 {\it{Exactly Solved Models in
Statistical Mechanics}} (Orland, FL: Academic)

\bibitem{bethe} H. Bethe Z. Phys. 71 (1932) 205.
\bibitem{fst} L. D. Faddeev, E.K. Sklyanin and L. A.
Takhtajian Theor. Math. Phys. 40 (1979) 194.
L.A. Takhtajian and L. D. Faddeev, Russ. Math. Surv. 34
(1979)11.
\bibitem{yy}C.N. Yang and C.P. Yang, Phys. Rev. 150 (1966)321,
327.
\bibitem{bt} L.A. Takhtajian Phys.Lett. A 87 (1982) 479.
H.M. Babujian, Nucl.Phys. B 215 (1983) 317.

\bibitem{KRS} P. Kulish, N. Reshetikhin and E. Sklyanin,
Lett. Math Phys. 5 (1981) 393-403.

\bibitem{KR}
A.N. Kirillov, N.Yu. Reshetikhin, J Phys. A: Math. Gen. 20
(1987), 1565-1585; and ibid. 1587-1597.

\bibitem{Jimbo}
M. Jimbo, Lett. Math. Phys. 10 (1985) 63-69.



\bibitem{KR81} P.D. Kulish and N. Yu. Reshetikhin,
Zap. Nauch. Semin. LOMI 101 (1981) 101,
translation in J. Sov. Math. 23 (1983) 2435.


\bibitem{Drinfeld} V.G. Drinfeld, Quantum groups in:
Proc. Intern Congress of Mathematics, Berkeley 1986,
I, pp. 798-820, California Acad. Press 1987.


\bibitem{FRT} L.D. Faddeev, N.Y. Reshetikhin and L.A. Takhtadjian,
in Algebraic Analysis, PP. 129-139, Acad. Prss, 1988

\bibitem{Woronowicz} S.L. Woronowicz, Commun. Math. Phys. 122 (1989)
125-170.

\bibitem{KST} V.N. Tolstoy and S,M. Khoroshkin,
Func. Anal. Appl. 26 (1992) 69-71;

S.M. Khoroshkin, A.A Stolin and V.N. Tolstoy,
Mod. Phys. Lett A10 (1995) 1375-1392.


\bibitem{BLZ} V.V. Bazhanov, S.L. Lukyanov and
A.B. Zamolodchikov, Commun. Math. Phys. 177 (1996) 381-398;
ibid. 190 (1997) 247-278; and ibid 200 (1999) 297-324.




\bibitem{Zhang}
G.~W.~Delius, M.~D.~Gould and Y.~Z.~Zhang,
Nucl.\ Phys.\ B {\bf 432} (1994) 377
[arXiv:hep-th/9405030].


\bibitem{BS}
V.V. Bazhanov and Yu.G. Stroganov, J. Stat. Phys. {\bf 59} (1990) 799.



\bibitem{LevPadua} L.N. Lipatov, {\it High-energy asymptotics of multicolor
QCD and exactly solvable lattice models} Padova preprint DFPD-93-TH-70B;
and
\newline
JETP Lett. B342 (1994)596.


\bibitem{FK}
L.~D.~Faddeev and G.~P.~Korchemsky,
Phys.\ Lett.\ B {\bf 342} (1995) 311
[arXiv:hep-th/9404173].






\bibitem{Lipatov:1993qn}
L.~N.~Lipatov,
Phys.\ Lett.\ B {\bf 309} (1993) 394.



\bibitem{Korchemsky:1995um}
G.~P.~Korchemsky,
Nucl.\ Phys.\ B {\bf 443} (1995) 255
[arXiv:hep-ph/9501232].



\bibitem{Korchemsky:1996be}
G.~P.~Korchemsky,
Nucl.\ Phys.\ B {\bf 462} (1996) 333
[arXiv:hep-th/9508025].








\bibitem{Lipatov:1999as}
L.~N.~Lipatov,
Nucl.\ Phys.\ B {\bf 548} (1999) 328
[arXiv:hep-ph/9812336].




\bibitem{DeVega:2001pu}
H.~J.~De Vega and L.~N.~Lipatov,
Phys. \ Rev. \ D{\bf 64} (2001) 114019,
arXiv:hep-ph/0107225.






\bibitem{Derkachov:2001yn}
S.~E.~Derkachov, G.~P.~Korchemsky and A.~N.~Manashov,
Nucl. \ Phys. \ B {\bf 617} (2001) 375,  arXiv:hep-th/0107193.







\bibitem{Karakhanian:2000gy}
D.~R.~Karakhanian and R.~Kirschner,
Fortsch.\ Phys.\  {\bf 48} (2000) 139
[arXiv:hep-th/9902031].





\bibitem{Derkachov:2000ze}
S.~E.~Derkachov, G.~P.~Korchemsky and A.~N.~Manashov,
Nucl.\ Phys.\ B {\bf 566} (2000) 203
[arXiv:hep-ph/9909539].





\bibitem{Braun:1998id}
V.~M.~Braun, S.~E.~Derkachov and A.~N.~Manashov,
Phys.\ Rev.\ Lett.\  {\bf 81} (1998) 2020
[arXiv:hep-ph/9805225].


\bibitem{Derkachov:2001km}
S.~E.~Derkachov and R.~Kirschner,
Phys.\ Rev.\ D {\bf 64} (2001) 074013
[arXiv:hep-ph/0101174].

\bibitem{Derkachov:2000ne}
S.~Derkachov, D.~Karakhanian and R.~Kirschner,
Nucl.\ Phys.\ B {\bf 583} (2000) 691
[arXiv:nlin.si/0003029].

\bibitem{Derkachov:2001sx}
S.~E.~Derkachov, D.~Karakhanyan and R.~Kirschner,
Nucl.\ Phys.\ B {\bf  618} (2001) 589,\
arXiv:nlin.si/0102024.

\bibitem{KS80} P.P. Kulish and E.K. Sklyanin, Zap. Nauchn. Semin. LOMI 95
(1980) 129.

\bibitem{TTF} V.O. Tarasov, L.A. Takhtadjian and L.D. Faddeev, Theor. Math.
Phys. 57 (1983) 163-181.

\bibitem{Nankai} E.K. Sklyanin, "Quantum Inverse Scattering Method", in
{\sl Quantum Groups and Quantum Integrable Systems}, (Nankai lectures),
ed. Mo-Lin Ge, pp. 63-97,
World Scientific Publ., Singapore 1992, [hep-th/9211111]

\bibitem{LesHouches} L.D. Faddeev, Les Houches lectures 1995,
hep-th/9605187.

\bibitem{Rahman} G. Gasper and M. Rahman, {\sl Basic hypergeometric series},
Cambridge University Press, 1990.

\end{thebibliography}
\end{document}